# Single MoS$_2$-flake as a high TCR non-cryogenic bolometer

Saba M. Khan[1,2], Jyoti Saini[1], Anirban Kundu[1], Renu Rani[1] and Kiran S. Hazra[1,*]

[1] Institute of Nano Science and Technology, Knowledge City, Sector 81, Mohali, Punjab 140306, India

[2] Department of Physics, Lancaster University, Lancaster, LA1 4YB, United Kingdom

[*]Corresponding email: kiran@inst.ac.in

**Abstract**

Temperature coefficient of resistance (TCR) of a bolometer can be tuned by modifying the thermal conductance of an absorbing materials since they sense radiations via the temperature change in the absorber. However, the thermal conductance of the absorber can be reduced by engineering the appropriate thermal isolation, which can be an ultimate solution towards making a highly sensitive thermal detector. Here, we have developed an atomically thin 2D bolometer detector made up of a mechanically transferred suspended multilayer-MoS$_2$ flake, eliminating the use of challenging thin-film fabrication process. The strength of our detector lies on the two factors: its large surface-to-volume window to absorb the radiations; the suspended configuration which prevents the heat dissipation through the substrate and therefore reduces the thermal conductance. The bolometric response of the detector is tested in both modes, via the photoresponse and the thermal response. The prototype is found to exhibit a very high TCR ~ -9.5%/K with the least achievable thermal noise-equivalent power (NEP) ~ 0.61 pWHz$^{-1/2}$, in ambient conditions at 328 K.

**Keywords**: Bolometers, Temperature co-efficient of resistance (TCR), Thermal conductance, 2D materials, IR detection, MoS$_2$

**Introduction**

Non-cryogenic bolometers are highly demanded due to their evolving applications in defense, security, surveillance, astronomy, tomography and smart energy systems,[1-6] which requires a temperature sensing material with a high temperature coefficient of resistance (TCR) and low 1/f noise properties without requiring any cooling module. However, for the room-temperature bolometers, achieving high TCR with a substantial signal-to-noise ratio has always been a challenging task.[7-10] In addition, the geometry of the absorber material is a crucial parameter that suffers major technological barriers for thin film based bolometer where fragile amorphous films are deposited in multilayer stacking.[11,12] However, after the emergence of 2D layered materials, new possibilities have emerged where the active material itself can act as an absorber due to their large aspect ratio and single-crystallinity. These materials can mechanically be transferred and suspended between pre-patterned electrodes in a free-standing stable condition to achieve the desired thermal isolation.[13,14] However, relentless efforts are made to develop non-cryogenic bolometers attaining high TCR based on graphene derivatives,[15-17] but the ultimate limitation with graphene is its metallic nature and vanishingly small electron-phonon coupling at room temperature which restraint the heat dissipation paths for electrons.[18-20] In this context, $MoS_2$ may prove to be an excellent candidate because of its semiconducting nature where the band gap is tuned with its layer numbers,[21] and also offers the carriers concentration dependent electronic and optical properties.[22] Besides, its phonon wavelength is compatible with the electronic transitions in momentum space, providing a superior electron-phonon coupling compared to other 2D materials.[23] Although, the electron-phonon coupling can further be enhanced by selecting a multilayer $MoS_2$ flake and becomes more robust with the increase in layer numbers via indirect band transition.[24,25]

We here demonstrate the bolometric effect in a single multilayer $MoS_2$ flake in ambient conditions, where a decent TCR with significantly low thermal noise-equivalent power (NEP) has been achieved. The atomically thin bolometer prototype exhibits fairly high responsivity upon IR illumination with ultrashort response time, where thermal isolation is achieved by engineering a suspended structure.[13] Our findings indicate that the massive absorption enabled by the large aspect ratio and the reduced thermal conductance due to suspended geometry are the key attributes in obtaining the profound bolometric response in the present configuration of the 2D $MoS_2$ bolometer. In addition, the thermal conductivity of the suspended MoS$_2$ flake is extracted utilizing the bolometric technique; putting forward a non-destructive technique to study the temperature-dependent transport properties in MoS$_2$-based devices.

**Results and Discussion**

The AFM height image of the suspended MoS$_2$ device is shown in **Figure 1a**, which is fabricated by mechanically transferring a MoS$_2$ flake on to the pre-patterned two gold electrodes on a Si/SiO$_2$ substrate. The inset is showing the layer thickness along the white dotted line in the main figure, the thickness of the flake is found to be ~ 103 nm which gives an idea of approximate layer numbers of ~150 when consider the interlayer spacing of ~ 0.6-1 nm.[26,27] The multilayer behavior of the flake can also be verified with its Raman spectra (**Figure 1b**), obtained at ambient conditions by utilizing the 532 nm laser line with a laser spot of ~1µm. The two Raman active modes, $E^1_{2g}$ at ~383.4 cm$^{-1}$ and $A_{1g}$ at ~408.5 cm$^{-1}$ appear in the spectra, which corresponds to in-plane Mo-S opposite vibration and out-of-plane S-S relative vibration, respectively. The difference between these two peak positions is found to be ~25.1 cm$^{-1}$, which depicts its multilayer arrangement as reported previously.[28]

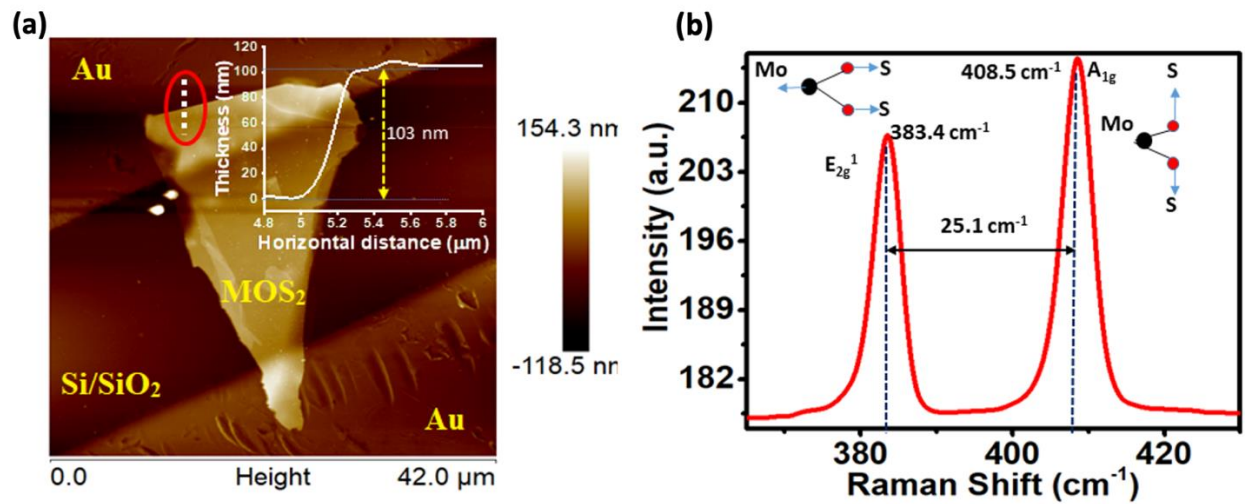

*Figure 1. (a) AFM height image of the as-fabricated device; the thickness of the MoS$_2$ flake is found to be ~103 nm which confirms a multilayer stacking of MoS$_2$. (b) Raman spectra of the MoS$_2$ contains both the characteristic Raman peaks corresponding to in-plane and out-of-plane modes, with the expected separation of ~ 25cm$^{-1}$ in the multilayer stacks.*

In order to probe the bolometric response of the suspended MoS$_2$ flake, an indigenous optical arrangement has been designed with a monochromatic light source with spectral resolution ~ 0.4 nm and an optical chopper. **Figure S1** (Supplementary Information) shows the schematic of the experimental set-up used to perform the measurements, where the electrical response in the device upon light irradiation of tunable wavelengths and powers are obtained using a standard Keithley source meter. A single NIR line of 900 nm (~ 1.38 eV) with ~ 5 µW is irradiated on the device, while the excitation energy ~ 1.38eV is deliberately kept lower than the optical band gap of the MoS$_2$ to avoid the photoconductive effect.[21,26,29,30] Furthermore, the possibility of photo-response from the thermoelectric effect is also eliminated by uniform exposure of NIR radiation on the whole device area. The chopping frequency is kept as low as 40 Hz to get the longer exposer time of NIR radiations onto the sample and thus achieving the saturation current to determine the sensitivity (S) and response time (τ) of the detector.

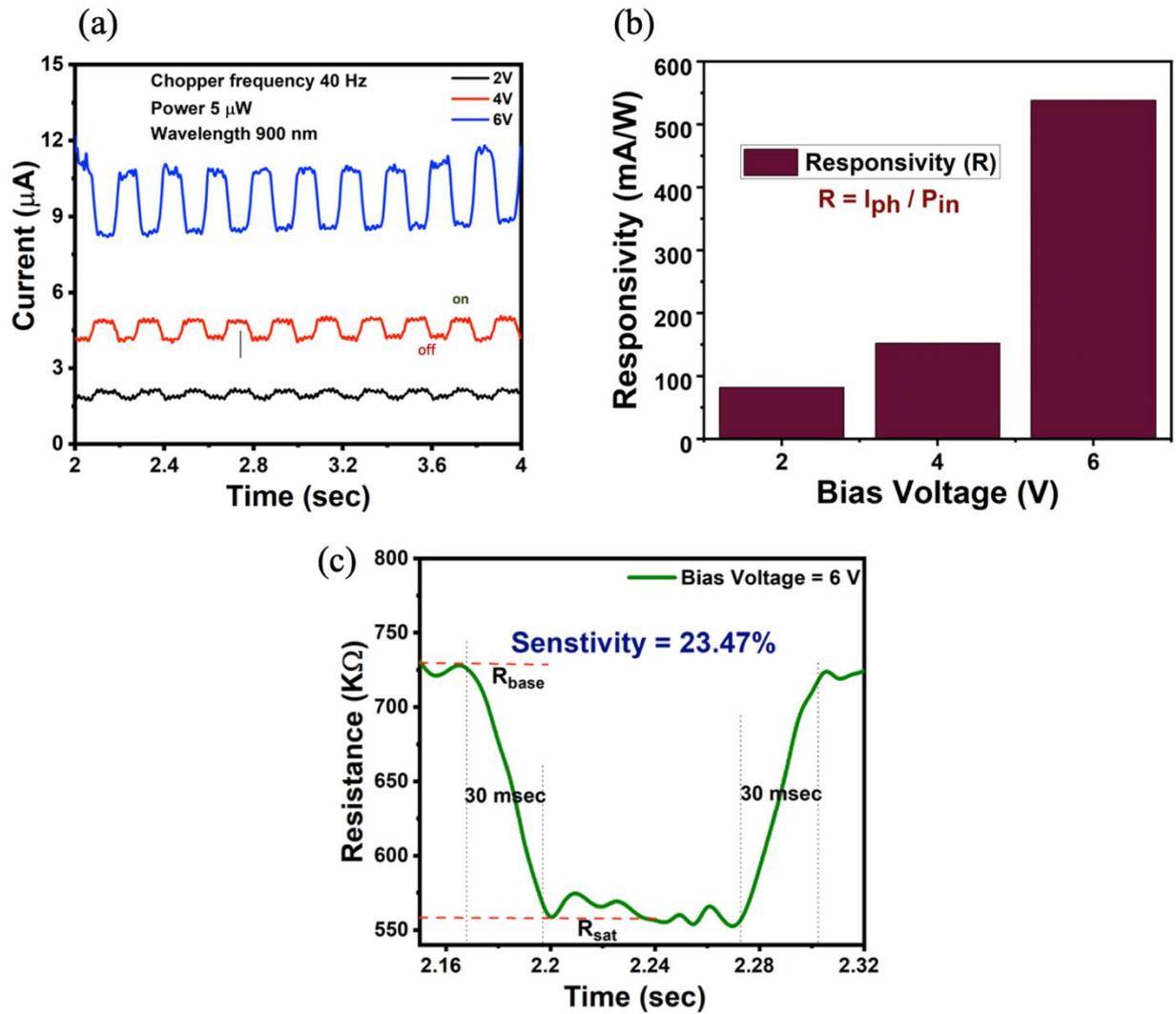

*Figure 2. (a) IR response of MoS$_2$ device for three constant bias-voltages at 40Hz chopping frequency. The detector shows a sharp increase in the current when exposed to radiations, whilst the current is rising with the applied bias. (b) Bias voltage dependence of responsivity for the suspended detector; sees a ~ 7 folds rise in the responsivity when a bias voltage is increased from 2V to 6V. (c) Resistance variation of the MoS$_2$ on illuminating with an IR beam at constant bias voltage of 6V, presents the response time of the device with baseline resistance and saturated resistance. The device exhibits a very short response and recovery times (~ 30 ms) with fairly high sensitivity.*

The change in electrical response with the applied light pulses of 40 Hz (pulse width ~ 25 msec) has been depicted in **Figure 2a**, which shows a sharp change in the channel current upon light irradiation. To improve the signal-to-noise ratio (SNR), the change in electrical response have been recorded with different applied bias, i.e., 2, 4 and 6 V, where the substantial increment is observed at 6 V. The increase in applied bias clearly improves the SNR (**Figure 2a**), which shows that at a constant bias of 6 V, a sharp and distinguishable change in electrical response is obtained. **Figure 2a** depicts the generation of periodic pulses of change in current with identical periodicity upon switching on the incident NIR pulses on the device. Upon switching on the NIR pulse, the current enhances rapidly and achieves the saturation followed by a sharp drop when switched off. The plot shows that the baseline and the saturation line are almost parallel, indicating that the changes were solely due to incident radiation.[31,32] The current responsivity (R) of the device has been determined using the equation, $R = \frac{I_{ph}}{P_{in}}$, where $I_{ph}$ and $P_{in}$ are the current under illumination and incident power, respectively. Responsivity of the device has been measured with varying applied bias (**Figure 2b**), which shows a significant improvement in the responsivity with increasing the applied bias. For an instance, the responsivity ~ 538 mA/W is obtained at applied bias 6V, which shows a ~ 7-fold enhancement as compare to the responsivity (~ 78 mA/W) at 2 V. Such higher responsivity of the device indicates its potential for using the micro-bolometer even at very low radiation power. Furthermore, the sensitivity (S) of the micro-bolometer device has been measured using the equation, $S = \frac{R_{base} - R_{sat}}{R_{base}} \times 100\%$, where $R_{base}$ and $R_{sat}$ are the base and saturation resistances, respectively. The change in resistance upon single pulse light irradiation at an applied bias 6 V is plotted in **Figure 2c**, which clearly depicts that the resistance sharply drops from its initial value ~ 729 kΩ at dark condition to ~ 558 kΩ upon light irradiation. Such sharp change in resistance offers a sensitivity (S) of ~ 23.47% at room temperature with ~ 5 μW power, which confirms the resolving capability with lower temperature fluctuation of the suspended MoS$_2$ at non-cryogenic temperature. The response time (τ) is determined by considering the time needed to change from 10% to 90% decrease in resistance, which is defined as $\tau = \frac{C}{G}$, where $C$ is the specific heat capacity of material and $G$ is thermal conductance. In **Figure 2c**, both the growth and decay response time has been obtained as ~ 30 msec, which is much slower as compared to the previously reported response time originated due to photocurrent generation, which confirms the bolometric response in the device.[10]

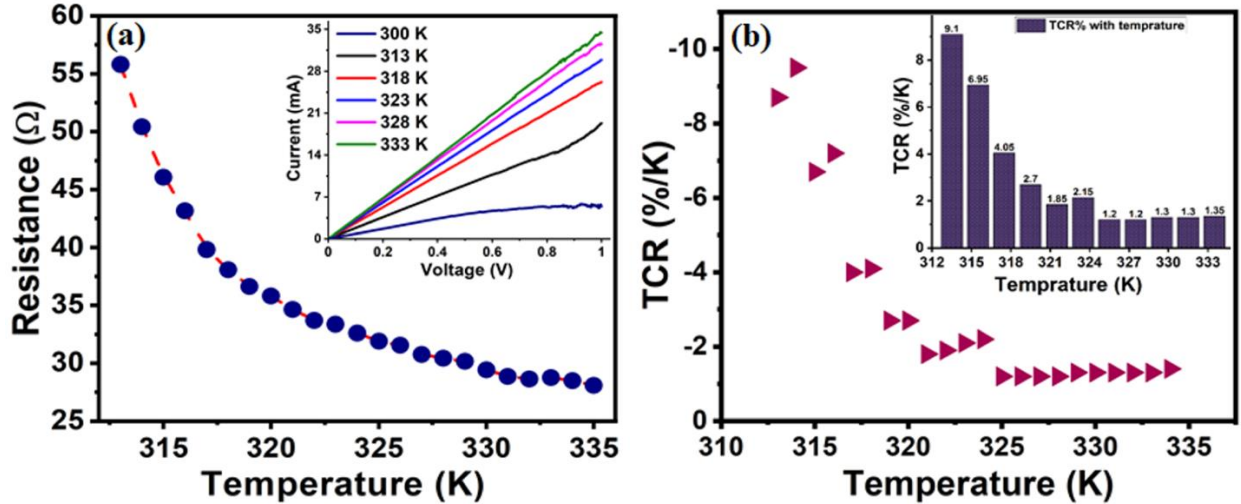

*Figure 3. (a) Resistance variation for the temperature range from 312K to 335K: The resistance of the device starts to decrease as we increase its temperature, and such a sharp variation in resistance with the temperature is a measure of the highly sensitive bolometric detection. **Inset**: I-V characteristics at different temperatures where current through the detector is increasing rapidly with the increase in temperature. (b) The percentage TCR against the temperature change: The device TCR is extracted via Eq.4 and is plotted as a function of temperature. Initially, the TCR is seeing a sharp decline with the temperature rise, but it starts to saturate at beyond 325K. Inset: the histogram of measured TCR versus device temperature.*

Although we have eliminated the possibility of inter-band transition by irradiation with lower energy wavelength than the band gap, yet to further confirm the bolometric response, the device is heated isothermally on a hot plate and corresponding electrical response is recorded to investigate the temperature response of the micro-bolometer (the suspended $MoS_2$ flake). In **Figure 3a**, we have plotted the change in resistance of the micro-bolometer with the increase in temperature, which shows exponential decay in resistance with the increase in temperature. The I-V responses with varying temperature are recorded by using two-probe method at normal atmospheric pressure with the change in substrate temperature, recorded by a J-type thermocouple having a swept from 40 °C to 62 °C in step of 1 °C (**inset of Figure 3a**). With the increase in temperature of the micro-bolometer device, a gradual rise in conductivity is observed which is originated due to the semiconducting nature of the $MoS_2$ flake.[33] However, the heat sensitivity of the suspended flake can be directly estimated by extracting its temperature coefficient of resistance

(TCR), which is a key figure of merit of a micro-bolometer since it determines the rate of change in resistivity with temperature. In normalized form, at a particular temperature the $TCR$ is defined as

$$\text{TCR}(\alpha) = \frac{1}{R}\left(\frac{dR}{dT}\right) \quad (1)$$

Which predicts that the materials having high $TCR$ offer large sensitivity ($S$) and so are suitable for bolometric application[34]. Using equation 1, the $TCR$ is plotted in **Figure 3b** at non-cryogenic temperatures varying from 313K to 333K. The maximum $TCR$ value is found to be ~ -9.5%/K at 314 K, which is significantly higher than that of the commercially available state-of-art non-cryogenic bolometers within the range of 0.01–3%/K.[35-40] Such high $TCR$ of the device can be attributed due to the large effective surface area of the MoS$_2$. In earlier reports,[31,34,41] it is anticipated that the improvement in the $TCR$ at cryogenic temperature can further be demonstrated by fabricating the cavity-like structure due to the enhanced infrared absorption, therefore, the suspended MoS$_2$ flake as a micro-bolometer may pave a path towards the development of next-generation bolometric sensors at non-cryogenic temperature.

In order to understand the mechanism behind the enhancement in current upon NIR irradiation, firstly the possibility of thermoelectric effect can be discarded as the experimental set-up provides uniform illumination/heating of the sample. The charge carriers (electrons in this case) gain some thermal energy from this uniform heating process during NIR illumination, which would equally be distributed among the charge carriers. Such uniform enhancement in the electron temperature leads to the reduction in the mean free path, and consequently a significant drop in the materials resistivity is observed. We have attributed this profound change in the resistivity to the suspended geometry of our device, where the electrons cannot dissipate their thermal energy towards the substrate which result in the increment of channel resistance.

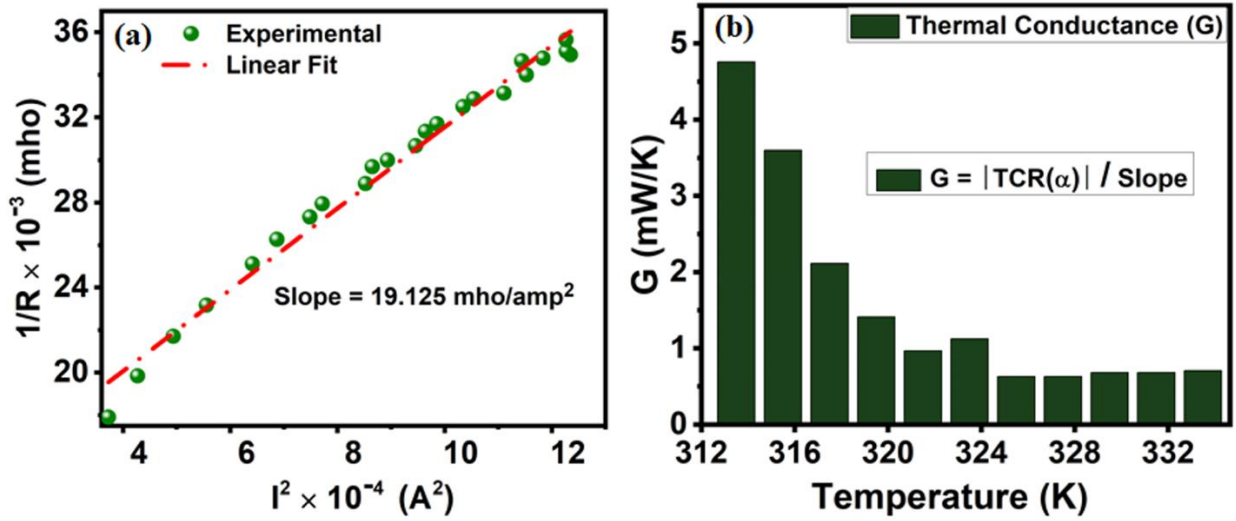

*Figure 4. (a) The 1/R against the square of the current is plotted, where the slope of the fit provides the TCR to thermal conductance ratio using equation 6. (b) The thermal conductance of the MoS2 device is extracted from the slope of (a) and is presented as a function of temperature. At first, the device conductance is reducing as temperature rises and then starts to plateau at 325K.*

The bolometric response of a device has a direct dependency on the thermal conductance of the active materials, i.e., $G = \frac{dP}{dT}$, or equivalent thermal resistance $R_h = dT/dP$, which quantifies the thermal response of the material upon incident radiation power ($P$). [41] To characterize the quantitative thermal conductance (or thermal resistance), we have assumed that the heat conduction is only possible through the supporting gold electrodes [42] for corresponding heat loss due to the Joule heating. Thus, the corresponding dissipated power will be equivalent to the thermal loss and can be defined as-

$$I^2 R = G(T - T_o) \qquad (2)$$

Now, comparing the (T-T$_O$) or dT values in equations (1) and (2), we have obtained

$$\frac{1}{R} = \frac{1}{Ro} - \left(\frac{\alpha}{G}\right) I_2 \qquad (3)$$

The Eq.3 implies that $1/R$ varies linearly with the $I^2$, and the slope of the fitted curve can be used to derive $\alpha/G$ ratio. The slope of the plot in **Figure 4a** curve has been used to extract the behavior of thermal conductance $G$ of the device with temperature (**Figure 4b**), where the $TCR\ (\alpha)$ values are taken from **Figure 3b**. **Figure 4b** shows the behavior of thermal conductance of the bolometer with temperature change; the thermal conductance associated with the maximum $TCR$ at 314 K is found to be ~ 5 mW/K, which has been reduced rapidly to ~ 0.63 mW/K at 328 K. By knowing the values of response time τ and $G$, one may easily find the specific heat capacity C of the multilayer flake using equation $\tau = \frac{C}{G}$, which is ~ 1.88 ×10⁻⁶ Joule per Kelvin at 328 K. Thermal and electrical properties of MoS$_2$ can be studied by analyzing the joule heating mechanism via the thermal conductance in the suspended bolometer, which is a major contribution of this work. Here, we introduce a novel technique to measure the thermal conductivity ($\kappa$) of the MoS$_2$ flake by utilizing the bolometric effect instead of Raman thermometry which has limited scope for the materials with less sensitive Raman peaks to the temperature variations. As we know thermal conductance of any nanomaterial is given by $G = A\kappa/(dx)$, where $A$ is the area of the nanomaterial, $\kappa$ is the thermal conductivity and $dx$ is thickness of the nanomaterial. In our device thickness ($dx$) of MoS$_2$ flake is ~ 103 nm and calculated area is ~ 30 μm². We can calculate the thermal conductivity ($\kappa$) if the thermal conductance ($G$) of nanomaterial is known. Using the above calculated thermal conductance of MoS$_2$ flake via bolometric effect, the extracted thermal conductivity at 314 K is ~ 17.17 W/mK, which is lower as compared to the previously reported thermal conductivity of the suspended few-layer MoS$_2$ flake.[43] This discrepancy can be explained in terms of the variation in the thermal conductivity with the MoS$_2$ flake thickness, the thermal conductivity of MoS$_2$ decreases as we increase the layer numbers.[44] Therefore, this single device can be used to extract all the transport properties in the multilayer MoS$_2$.

Following the thermal conductance measurement, the signal-to-noise ratio of the detector is estimated which is a measure of the detectivity of a detector. However, the high signal-to-noise ratio of bolometer detectors yields low noise-equivalent power ($NEP$). In a measurement set-up, many noise sources could be present such as those from the amplifiers,[34,45,46] but some are inherent to the bolometer itself namely the Johnson–Nyquist noise and the thermal fluctuations.[34] The Johnson–Nyquist noise can be determined from the expression $\frac{\sqrt{\frac{4kBT}{R}}}{R_{res}}$, which turns out to be ~ 43.04 pWHz⁻¹ᐟ² for $R_{res}$ 538 mA/W at 328 K. However, the thermal fluctuation $NEP$ directly depends on the thermal conductance ($G$) and the temperature ($T$) of the bolometer and is defined as $\sqrt{(4kBT^2G)}$.[45] At 328 K, the $NEP$ due to thermal nose is ~ 0.61 pW Hz⁻¹ᐟ² corresponding to measured $G$ ~ 0.63 mW/K. Therefore, the measured $NEP$ of MoS$_2$ microbolometer from the intrinsic noise sources is remarkably low in-contrast to the so far existing non-cryogenic bolometers[16], this signifies the novelty of our detector with its large signal-to-noise ratio. The intrinsic $NEP$ can further be improved by reducing the thermal loss of the IR absorbing material and by increasing the responsivity. The **Table 1** given below shows the comparison of $TCR$ of our suspended MoS$_2$ device with the other previously reported devices.

**Table 1.** TCR values of available non-cryogenic bolometer detectors

| Sl. No. | Detector Type | TCR value | Reference |
|---|---|---|---|
| 1 | Graphene | ≈ 11% /K at 296 K | 10 |
| 2 | Reduced Graphene Oxide | ≈ 44×10$^3$ Ω/K at 60K | 38 |
| 3 | Ti | ≈ 0.27% /K | 44 |
| 4 | Graphene Aerogel | ≈ 0.3% /K at 295 K, ≈ 2.9%/K at 10 K | 15 |
| 5 | SWCNT | ≈ 2.94% /K at 300 K | 45 |
| 6 | VOx | ≈ 2.1%/K at 298 K | 46 |
| 7 | MoS$_2$ | ≈ 9.5% /K at 314 K | Present Work |

Furthermore, we have developed a single pixel prototype of thermal sensor (designed on the Arduino platform) based on an equivalent suspended MoS$_2$ micro-bolometer, as shown in the circuit diagram of **Figure 5a**. User-contributed function libraries like PWM control, I2C and SPI are used to program the Arduino board.

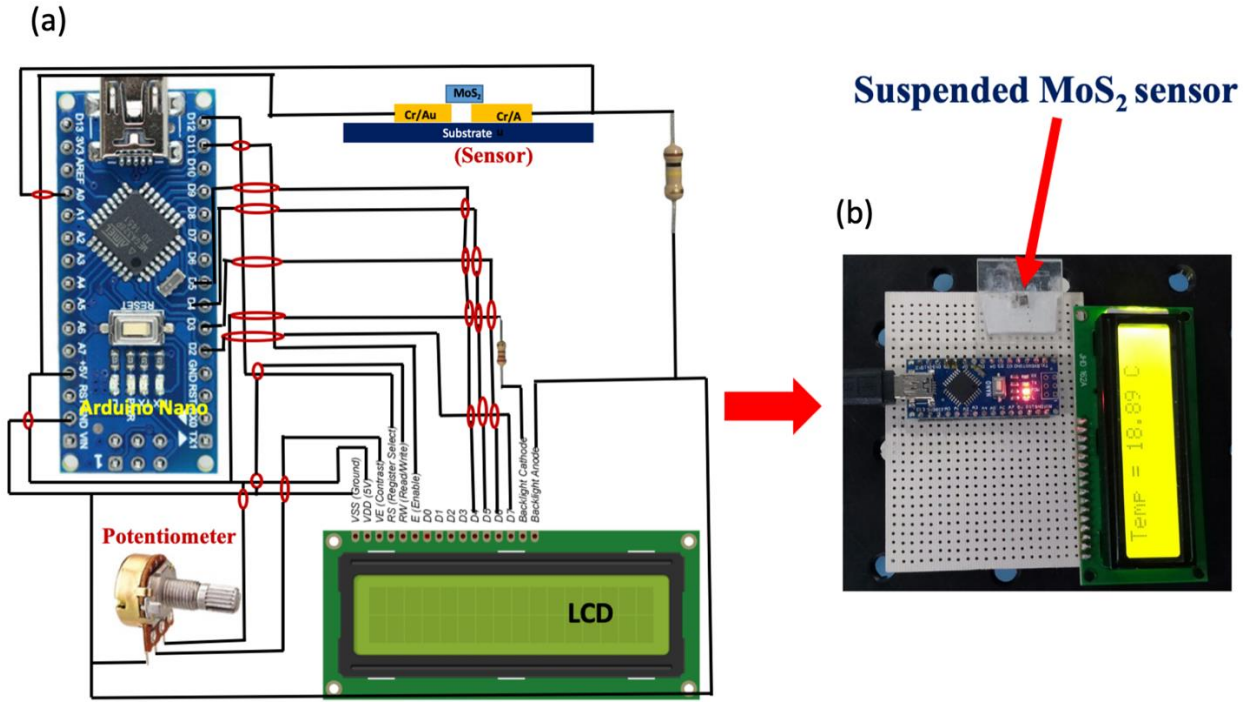

*Figure 5.* (*a*) *Circuit diagram of the suspended MoS$_2$ flake-based prototype bolometer, designed on the Arduino platform where an LCD is used to directly demonstrate the temperature variation of the detector upon IR irradiation.* (*b*) *Real image of our prototype assembled on a breadboard: the detector is mounted on a glass slide and soldered to a Nano Arduino along with a LCD.*

During the IR irradiation on the MoS$_2$ micro-bolometer, the temperature variation caused by the radiation exposure led to the change in the resistance of the sensor which is recorded through the digital pins of the Arduino and matches with the pre-calibrated database of the current response with temperature. Microcontroller board of the Arduino then processes these signals as per the given coding and sends them to the sensor's output through its analogue pin to display it on the LCD screen, as demonstrated in **Figure 5b**. Hence, the suspended multilayer MoS$_2$, which is thermally isolated from the substrate to reduce the heat loss, can be a promising material for the future non-cryogenic bolometer technology concerned to the thermal detection.

**Conclusion**

We have investigated the non-cryogenic bolometric response in a suspended multilayer MoS$_2$-flake, where its free-standing architecture eliminates the heat exchange possibility with the substrate and results in the high-yield conversion of the absorbed energy into an electrical signal.

Such higher absorption capability leads to the maximum achievable $TCR$ of −9.5%/K at 314K, which is much higher than the commercially available Si or $VO_x$ based non-cryogenic bolometers. Furthermore, the thermal conductance G, which limits the speed of the bolometer, is found to be fairly low (~ 4.97 mW/K at 314K), and consequently, the device has least thermal $NEP$ of 0.61 pW $Hz^{-1/2}$ at 328K with the Johnson noise of ~ 43.04 pW $Hz^{-1/2}$. Finally, a prototype of single pixel temperature sensor has been proposed which can sense small energy differences like IR-radiations. Since the heat loss prevails through the gold electrodes of the suspended device, the optimized gold electrode's structure may elevate the $MoS_2$-based future bolometric devices. Also, the sensitivity of the sensor may further be upgraded by enhancing the absorption of the incident radiations by fabricating the micro (or) nano cavities.

**Experimental Details:**

The suspended $MoS_2$ devices are fabricated with photolithography by writing a rectangular pattern of electrodes on a 290 nm-thick $SiO_2$ substrate underlying the highly doped p-type Silicon. Subsequently, the electron beam deposition technique is used to deposit a stack of Cr/Au (10/60nm) on these pre-written electrodes. Then, the micromechanical exfoliation method is considered to transfer a multilayer $MoS_2$ flake onto the PDMS from a single crystal of $MoS_2$, and finally, the flake is transferred from the PDMS onto the two rectangular gold electrodes separated by 20µm gap using a home-built transfer stage.

**Supporting Information**

The Supporting Information includes the following details.

**Figure S1.** Photoluminescence (PL) spectra of $MoS_2$

**Figure S2.** I-V characteristic of $MoS_2$ device in dark indicating the ohmic behavior of the contact between flake and gold electrodes

**Figure S3.** Schematic diagram for $MoS_2$ device and measurement set-up

**Figure S4.** Band alignment of both suspended (a) and supported (b) $MoS_2$ device under IR illumination

**Author Information**

*Correspondence to be addressed to K.S.H (kiran@inst.ac.in)

Kiran Shankar Hazra - Institute of Nano Science and Technology, Knowledge City, Sector 81, Mohali, Punjab-140306, India

Email: kiran@inst.ac.in

**Author Contributions**

S.M.K. and K.S.H. designed the experiments and S.M.K. performed all the measurements. S.M.K., A.K. analysed the data and prepare the original draft. J.S., A.K. and R.R did the formal analysis and data curation; K.S.H. supervised the research project and corrected the manuscript.

**Acknowledgments**

K.S.H. would like to acknowledge the grant sanctioned by the DST-SERB under EEQ scheme (Project No. EEQ/2017/000497). S.M.K. and J.S. would like to acknowledge the DST-SERB for their fellowship through the project grant (EEQ/2017/000497). S.M.K. would like to thank Mamta Raturi for the technical assistance with the AFM data and Saif Ali Khan for the fruitful discussion and comments on the manuscript.

# Supporting Information

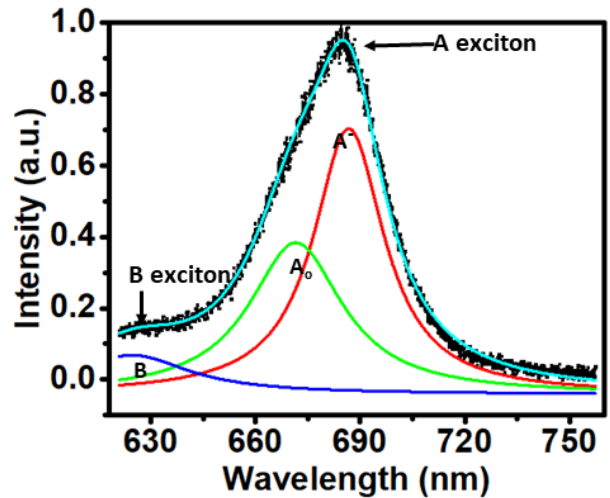

*Figure S1. Photoluminescence (PL) spectra of MoS$_2$*

The multilayer MoS$_2$ flake shows a very strong Photoluminescence (PL) peak centred around 689 nm with the asymmetric distribution (**Figure S1**). The Lorentzian peak fitting indicates the presence of the three distinct peaks at wavelengths 622.2 nm, 671.8 nm and 686.5 nm. The first two peaks are due to the existence of neutral excitons $B$ and $A_0$ respectively, near the $K$ point of the brillouin zone boundary[1,2]. These two excitons are expected to exist due to spin-orbit band splitting in the valence band and the interlayer coupling[3]. The third Lorentzian peak, observed at 686.5 nm, is associated with the trions (charged excitons), which is three particle system comprising of an exciton with an electron (or a hole), known as the negative (or positive) trion [4]. In PL spectra, negative trions ($T$) recombination is dominant which indicates the intrinsic n-type behaviour of MoS$_2$[4].

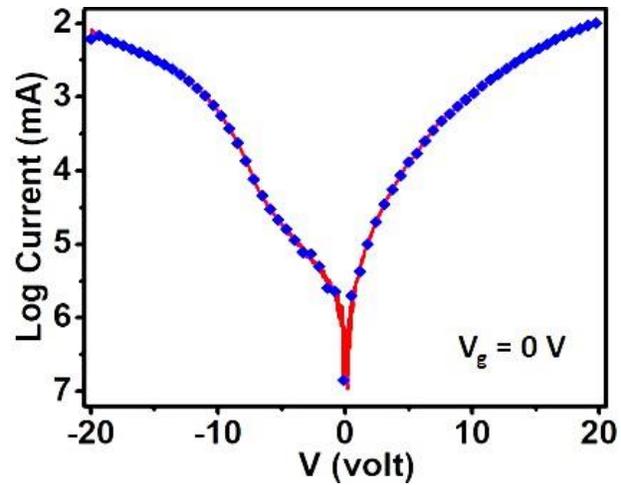

*Figure S2.* I-V characteristic of MoS$_2$ device in dark indicating the ohmic behaviour of the contact between flake and gold electrodes.

To investigate the nature of contact formed between the suspended flake and the electrodes, the output characteristics is plotted in semi-log scale for zero gate voltage, shown in **Figure S2.** The output current is almost identical and symmetric for both the forward and reverse bias, which indicate a good ohmic contact at the MoS$_2$-Au junction.

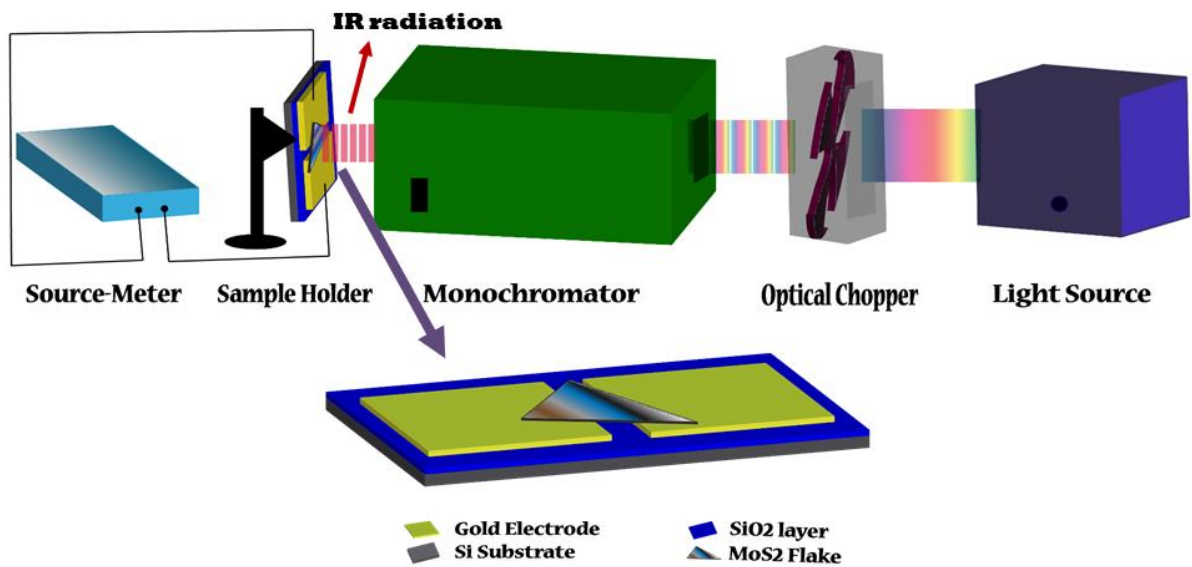

*Figure S3.* Schematic diagram for MoS$_2$ device and measurement set-up.

**Figure S3** shows a schematic representation of the MoS$_2$ device and experimental set-up to examine the photo response of the device. A white-light beam from the light source is entered into the monochromator via passing through the optical chopper, which chopped the light at the desired frequency. A 900 nm IR wavelength is selected from the monochromator, incident on the MoS$_2$ flake and then corresponding I-V is measured by the source-meter connected to the sample by through the gold electrodes. Here the monochromator is kept 10 cm away from the device in ambient conditions and the incident radiation power at the device position is measured by a Si based power meter.

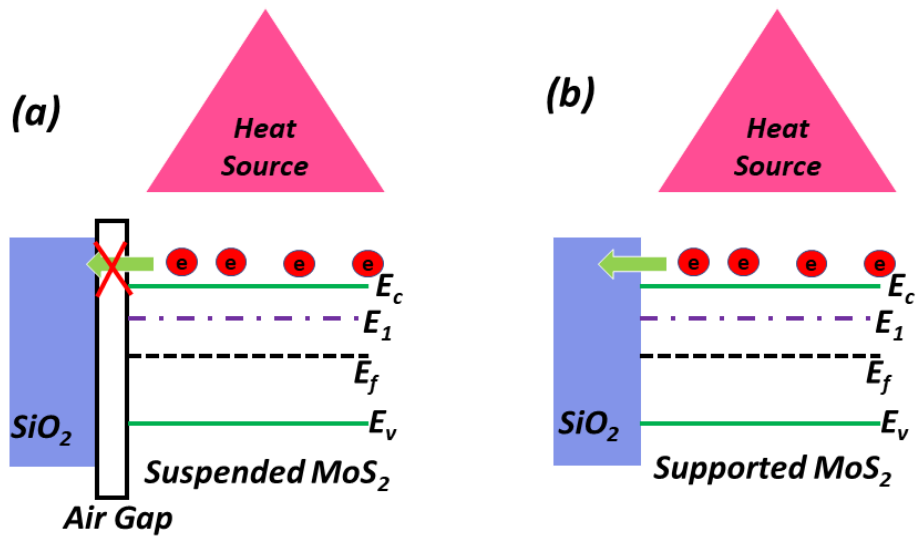

*Figure S4.* Band alignment of both suspended (**a**) and supported (**b**) MoS$_2$ device under IR illumination

To understand the mechanism behind the rise in the current upon NIR irradiation / heating, the band alignment of the device is sketched in **Figure S4.** Our experimental set up uniformly illuminates/warms-up the device so that there is no temperature gradient across the suspended MoS$_2$ flake and therefore, the possibility of thermoelectric effect can be ignored.

The charge carriers (electrons in the present case) gain some thermal energy from this heating process, which would equally be distributed among them. Such even rise in the temperature of the electrons lead to the reduction in the mean free path, and hence results in a significant drop in the material resistivity. We attribute this profound change in the resistivity with the suspended geometry of our device as electrons cannot dissipate their heat energy towards the substrate, and so are giving rise the resistance variation with change in the temperature of the MoS$_2$ flake.